\documentclass[12pt]{article}
\usepackage{graphicx} 
\usepackage{authblk}
\usepackage{amsfonts,amsmath,amssymb,epsf}
\usepackage{amsmath,braket}
\usepackage[dvipsnames,svgnames,table,x11names]{xcolor}
\usepackage{bm,comment,footnote}
\usepackage{float}
\usepackage[makeroom]{cancel}
\usepackage{color}
\usepackage{graphicx}
\usepackage{amsmath,braket}
\usepackage{amssymb}
\usepackage{xspace}
\usepackage[small]{subfigure}
\usepackage[numbers,sort]{natbib}
\usepackage[hyperfootnotes=false]{hyperref}
\usepackage[most]{tcolorbox}
\usepackage[capitalize]{cleveref}
\usepackage{hyperref}
\usepackage{framed}
\usepackage{physics}
\usepackage{tensor}
\usepackage{mciteplus} 
\usepackage{dcolumn}
\usepackage{bm}
\usepackage{verbatim}
\usepackage{amscd}
\usepackage{amsfonts}
\usepackage{setspace}
\usepackage{amsthm}
\usepackage{enumerate}
\usepackage{mathtools}
\usepackage{framed}
\usepackage{xcolor}
\usepackage{todonotes}
\usepackage{enumitem}

\newcommand{\currentproblemtarget}{}

\newcommand{\problemlistlabel}{%
    \hyperlink{\currentproblemtarget}{\arabic{enumi}.}%
}

\newlength{\fighskip} \fighskip=2pt
\newlength{\figvskip} \figvskip=3pt

\newcounter{probcount}[section]
\renewcommand{\theprobcount}{\thesection.\arabic{probcount}}

\newcommand{\problem}[1]{%
    \refstepcounter{probcount}%
    \hypertarget{prob-\thesection-\arabic{probcount}}{}%
    \begin{tcolorbox}[
        enhanced,
        breakable,
        colback=white,
        colframe=black!80,
        arc=2pt,
        outer arc=2pt,
        leftrule=3pt,
        fonttitle=\bfseries,
        title={Problem \theprobcount}
    ]
        #1
    \end{tcolorbox}
}

\newcommand{\problink}[2]{%
    \hyperlink{prob-#1-#2}{#1.#2}%
}

\newcommand{\problemrefs}[1]{%
    \if\relax\detokenize{#1}\relax
    \else
        \quad {\footnotesize$\to$ #1}%
    \fi
}

\title{
\vspace{-80pt}
\hfill
{\small RIKEN-iTHEMS-Report-26, OU-HET-1312, KEK-TH-2845, KUNS-3110, RUP-26-14, YITP-26-79}\\
\vspace{50pt}
\hspace{20pt}
\bf Rethinking quantum information \\ in gravity and fields
}

\author[1,2]{Kanato Goto}
\author[2,3,4]{Yuta Hamada}
\author[5]{Kohtaro Kato}
\author[6]{Takato Mori}
\author[7,8]{\\Yoshifumi Nakata}
\author[9]{Masataka Watanabe}
\author[10]{Hayata Yamasaki}
\author[9,11,12,13,2]{\\Masahito Yamazaki}
\author[14]{Takuya Yoda}

\affil[1]{Department of Physics, Graduate School of Science, The University of Osaka, \protect\\
Machikaneyama-Cho 1-1, Toyonaka 560-0043, Japan}
\affil[2]{RIKEN Center for Interdisciplinary Theoretical and Mathematical Sciences (iTHEMS), RIKEN, \protect\\
2-1 Hirosawa, Wako, Saitama 351-0198, Japan}
\affil[3]{Theory Center, IPNS, KEK, 1-1 Oho, Tsukuba, Ibaraki 305-0801, Japan}
\affil[4]{Graduate University for Advanced Studies (SOKENDAI), 1-1 Oho, Tsukuba, Ibaraki 305-0801, Japan}

\affil[5]{Department of Mathematical Informatics, Graduate School of Informatics, Nagoya University, 
\protect\\
Furo-cho, Chikusa-ku, Nagoya 464-0814, Japan}
\affil[6]{Department of Physics, Graduate School of Science, Rikkyo University,\protect\\
3-34-1 Nishi-Ikebukuro, Toshima-ku, Tokyo 171-8501, Japan}
\affil[7]{Department of Computer Science, School of Computing, 
Institute of Science Tokyo, \protect\\
4259 Nagatsuta-cho, Midori-ku, Yokohama, Kanagawa 226-8501, Japan}
\affil[8]{Yukawa Institute for Theoretical Physics, Kyoto University, \protect\\
Oiwake-cho, Kitashirakawa, Sakyo-ku, Kyoto, 606-8502, Japan}
\affil[9]{Department of Physics, Graduate School of Science, University of Tokyo, \protect\\
7-3-1 Hongo, Bunkyo-ku, Tokyo 113-0033, Japan}
\affil[10]{Department of Computer Science, Graduate School of Information Science and Technology, The University of Tokyo, 7-3-1 Hongo, Bunkyo-ku, Tokyo 113-0033, Japan}
\affil[11]{Kavli Institute for the Physics and Mathematics of the Universe (Kavli IPMU, WPI), UTIAS, University of Tokyo, 5-1-5 Kashiwanoha, Chiba 277-8583, Japan}
\affil[12]{Trans-Scale Quantum Science Institute, University of Tokyo, 7-3-1 Hongo, Bunkyo-ku, Tokyo 113-0033, Japan}
\affil[13]{Center for Data-Driven Discovery (CD3), Kavli Institute for the Physics and Mathematics of the Universe (Kavli IPMU, WPI), UTIAS, University of Tokyo, 5-1-5 Kashiwanoha, Chiba 277-8583, Japan}
\affil[14]{Department of Physics, Graduate School of Science, Kyoto University, \protect\\
Kitashirakawa, Sakyo, Kyoto 606-8502, Japan}

\date{}

\begin{document}

\maketitle
\thispagestyle{empty}

\begin{abstract}
This paper presents a curated selection of research questions at the intersection of quantum gravity and quantum information, chosen to highlight issues that we regard as particularly important for researchers in both fields. We organize the discussion into four main themes: the operational characterization of observables, the role of observers, quantum error correction, and the infinite-dimensionality of Hilbert spaces. We hope that addressing these questions will engage researchers across both communities and further strengthen the profound interplay between the two disciplines.
\end{abstract}

\newpage

\tableofcontents


\section{Introduction}

Historical precedents attest that major scientific breakthroughs often emerge at the intersection of seemingly disparate disciplines. As Werner Heisenberg observed, ``the most fruitful developments frequently take place at those points where two different lines of thought meet,'' a sentiment that accurately characterizes the current frontier of theoretical physics.

In the twenty-first century, we are witnessing the synthesis of two distinct ``quantum'' disciplines: quantum gravity (QG) and quantum information (QI). The confluence of these two vast intellectual streams is already yielding a novel language capable of describing the fundamental nature of the universe. 
It is fair to say, however, that significant effort is still needed to bridge the differences between the two disciplines, and there are reasons to believe that the culmination of this synthesis has yet to come.

In this report, coauthored by early-career researchers in both QG and QI, we list selected open problems at the intersection of the two disciplines. 

We have categorized the problems into the following four themes:\footnote{Naturally, some questions are relevant for discussions in multiple sections. We have chosen the category that best fits each problem.} 
\begin{itemize}
    \item Operational characterizations (\cref{sec:operational}),
    \item The role of the observer (\cref{sec:observer}),
    \item Quantum error correction (\cref{sec:QEC}),
    \item Infinite-dimensionality of Hilbert spaces (\cref{sec:infinite_dim}).
\end{itemize}
While this list is by no means comprehensive and inevitably reflects the authors' preferences and perspectives, we have selected particular problems that we believe will be of fundamental importance in further elucidating the connections between the two areas. 

Each section begins with a very brief explanation of the background and the overarching theme. We then discuss each problem in turn, providing more detailed explanations as we go. We have included both technical and conceptual questions. 
In Appendix \ref{sec:list} we have compiled a list of all problems; this is partly for the convenience of readers, and partly to highlight synergies between different problems.

Since the topics discussed in this report are diverse, the references are by no means exhaustive. Instead, they serve as entry points for readers interested in exploring specific areas in greater depth, and we have included both original references and review articles.
We apologize in advance for any significant omissions.

We hope that this report will help stimulate future developments in QG and QI.

\section{Operational characterization}\label{sec:operational}

Modern analyses of quantum field theory (QFT) and QG frequently invoke entropic quantities originating in QI.
While these quantities are often treated as mathematical functions assigning numbers to quantum states, QI goes further and endows them with \emph{operational meaning}, thereby shedding new light on their physical significance:\footnote{As humorously illustrated by the line ``The answer to the ultimate question of life, the universe, and everything is 42,'' from ``The Hitchhiker's Guide to the Galaxy,'' a number alone carries no value unless its meaning is specified.} entropic measures characterize the optimal performance of \emph{information-processing tasks} such as compression, transmission, state discrimination, and state conversions.
For example, the entanglement entropy of a bipartite pure state admits an operational interpretation via asymptotic entanglement distillation (i.e., converting the given state to Bell pairs) and dilution (i.e., forming the given state under local operations and classical communication (LOCC))~\cite{Bennett:1995tk}. 
Similarly, the mutual information is interpreted as the amount of noise required to completely erase the correlations in a given state~\cite{Groisman:2005dbo}.

In this section, we ask whether QG observables admit analogous operational characterizations, and conversely, whether information-processing tasks can be used to define new quantities relevant for holography (\cref{subsec:tasks}).
This perspective can be systematically formulated within the framework of resource theories (\cref{subsec:resource_theory}).

\subsection{What is the role of information-processing tasks in QG?}
\label{subsec:tasks}
Because QI is not typically formulated in settings with dynamical spacetime, it is natural to ask why information-processing tasks should be relevant to QG. 
Our motivation is that such tasks may provide a sharp probe into the nature of QG: they require us to specify what can actually be prepared, transmitted, reconstructed, or verified by physical observers subject to constraints such as causality, gauge invariance, and limited access. In this way, they do not merely assign new interpretations to preexisting quantities, but can justify (or even define) quantum gravitational observables through their operational meaning. More generally, even when a quantity is well defined and computable, it is often unclear what physical capability or limitation it captures.
A task-based perspective addresses this gap by relating such quantities to concrete operational possibilities.
From this viewpoint, information-processing tasks offer a perspective complementary to approaches based solely on formal observables or semiclassical geometry.\footnote{By semiclassical, we mean bulk effective field theory: quantum matter and perturbative metric fluctuations (gravitons) propagating on a classical or weakly fluctuating curved spacetime background.}

\newcommand{\textTwoOne}{Can we characterize gravitational observables as operational quantities on the boundary and/or via gravitational tasks?}
\problem{\textTwoOne}
While gravitational (i.e., diffeomorphism-invariant) observables, such as extremal surfaces/volumes and dressed bulk local operators, are well-defined in semiclassical gravity, it is nontrivial to determine whether they remain meaningful in full QG. A natural question is whether there exist information-processing tasks whose achievability or optimal performance is directly related to such observables.
If the task is defined independently of $G_N$ (Newton's constant), then finite-$G_N$ implementations may provide a consistent definition of \emph{quantum} gravitational observables.\footnote{Finite $G_N$ means that we consider a regime beyond the strict $G_N\rightarrow 0$ limit, which is governed by semiclassical gravity described by general relativity and quantum field theory on curved spacetime.}

In the context of AdS/CFT,\footnote{Roughly speaking, the holographic principle suggests that a theory of quantum gravity in a bulk spacetime can be equivalently described by a non-gravitational quantum theory living on its boundary. AdS/CFT provides a concrete realization of this idea, where the boundary theory is often a large-\(N\) strongly coupled quantum field theory with \(O(N^2)\) degrees of freedom.} there has been substantial progress in relating bulk quantities to boundary information-theoretic quantities. For instance, the area of a bulk extremal surface~$\gamma_A$ bounding a region $A$ is related to the entanglement entropy $S_A$ of the dual conformal field theory (CFT)~\cite{Ryu:2006bv,Ryu:2006ef,Hubeny:2007xt,Engelhardt:2014gca}. However, this relation is not originally derived from an explicit operational task, but rather from the identification of $S_A$ with the von Neumann entropy $S_A=-\tr\rho_A\log\rho_A$, where $\rho_A$ is the reduced density matrix of a subsystem of interest. 
Since entanglement entropy of pure states admits an operational interpretation via LOCC entanglement distillation and dilution in the i.i.d.\ limit, namely, in the limit of a large number of copies of the original state,\footnote{See standard textbooks such as Refs.~\cite{Nielsen:2012yss,Wilde:2013sha} for a precise definition.} it is natural to ask whether a similar operational interpretation exists in holography. 

Concretely, let us restrict attention to holographic states, by which we mean boundary states that admit a semiclassical bulk description around some gravitational saddle. In this setting, the relevant low-energy bulk observables are described within a code subspace\footnote{A term borrowed from quantum error correction. See Problem~\problink{4}{4} for a brief explanation.} associated with that saddle. More generally, different semiclassical saddles may give rise to different effective Hilbert-space descriptions, for instance via Gelfand-Naimark-Segal (GNS) constructions around the corresponding semiclassical states~\cite{Liu:2025krl}. Thus, the holographic CFT Hilbert space should not be identified with a single bulk Hilbert space; rather, semiclassical bulk physics is captured by suitable code subspaces or low-energy operator algebras associated with particular saddles.

One may then restrict not only the states but also the allowed operations. A natural class consists of operations that preserve the relevant semiclassical sector, or more generally, map one semiclassical holographic sector to another. Low-energy operations such as single-trace excitations are examples, while sufficiently controlled high-energy operations may backreact on the geometry and map the state to another state corresponding to a different semiclassical saddle. In tensor-network models of holography, this restriction can be made more concrete by considering operations that map tensor-network states within a fixed network/geometry ansatz~\cite{Pastawski:2015qua}.

With these restrictions in mind, one can ask whether the entanglement of a holographic bipartite pure state $\ket{\Psi_{\rm hol}}_{AB}$ can be distilled into Bell pairs by a correspondingly restricted class of LOCC, which we loosely refer to as holographic LOCC. Namely, does the relation
\begin{equation}
    \Lambda_{AB}(\ket{\Psi_{\rm hol}}_{AB}) \stackrel{?}{\approx} \ket{\rm EPR}^{\otimes \frac{\mathrm{Area}(\gamma_A)}{4G_N}}, \quad \Lambda\in\mathrm{holographic\ LOCC}
    \label{eq:hol-dist-q}
\end{equation}
hold up to $O(1)$ corrections, where $\ket{\rm EPR}$ is a maximally entangled state?\footnote{Note that the bulk procedure for entanglement distillation, implemented via entanglement wedge reconstruction, acts on a single copy of the state. Thus, the relevant setting is one-shot rather than asymptotic. A careful reader may therefore wonder why \eqref{eq:hol-dist-q} is expected without taking the i.i.d. limit. Strictly speaking, one should formulate the statement in terms of smooth one-shot entropies~\cite{Akers:2023fqr}. However, in the semiclassical limit $G_N\to 0$, the min- and max-entropies coincide at leading order, in a manner analogous to the i.i.d. limit~\cite{Bao:2018pvs}. Hence, one expects \eqref{eq:hol-dist-q} to hold up to small errors with the von Neumann entropy.}
Using entanglement wedge reconstruction (EWR)~\cite{Cotler:2017erl,Chen:2019gbt}, one can interpret a one-shot distillation task as `pushing the boundary' toward the extremal surface via local unitaries removing encoding isometries~\cite{Mori:2024gwe,Mori:2022xec,Bao:2023til,Lin:2020yzf}.\footnote{The operational interpretation of entanglement wedge reconstruction is subtle and not unique. While it is originally motivated by the JLMS formula~\cite{Jafferis:2015del}, showing the perturbative equality between the boundary and bulk relative entropies, its precise task interpretation requires careful analysis. In particular, \cite{Akers:2020pmf} has pointed out that there exists a unitary operator that disentangles bulk entanglement and obstructs certain bulk operators in the original entanglement wedge. This implies that reconstructability is a state-dependent statement. To correctly understand what we really mean by the EWR, it should be formulated as a task with quantitative thresholds with specified states rather than a binary criterion. Alternative interpretations include state merging, a more general information-transfer task that includes entanglement distillation as its subset~\cite{Bousso:2023sya}.
}
In tensor-network models, this procedure identifies Bell pairs across the extremal surface and establishes
\begin{equation}
    \ket{\Psi_{\rm hol}}_{AB} \approx ( V_A \otimes W_B ) \ket{\rm EPR}^{\otimes \frac{\mathrm{Area}(\gamma_A)}{4G_N}}, \quad V^\dag V=\mathbf{1}_A,\ W^\dag W=\mathbf{1}_B.
\end{equation}
By extending such one-shot tasks to holographic mixed states, one finds that another type of minimal-area surface called the entanglement wedge cross section governs the distillable entanglement~\cite{Mori:2024gwe}. In addition, certain LOCC distillation protocols admit gravitational realizations, e.g. traversable wormholes~\cite{Gao:2016bin,Maldacena:2017axo,Susskind:2017nto,Wang:2024yqn}.

While the interpretation of the bulk-to-boundary map as entanglement distillation provides a relatively direct operational characterization of the holographic entanglement entropy, many gravitational observables are not immediately captured by such tasks. It is therefore important to understand how to operationally characterize these observables that are not directly associated with standard information-processing tasks. For example, an angle formed by two minimal-area surfaces is related, under suitable conditions, to modular commutators in the dual CFT~\cite{Zou:2022nuj}. 
This suggests that the angle might be operationally characterized by tasks whose performance is governed by noncommutativity, such as Clauser–Horne–Shimony–Holt (CHSH) games~\cite{Clauser:1969ny,Cleve:2010xmq}.

Some geometric quantities even lack a clear boundary definition in terms of density matrices. For instance, the minimal tri-way cut area, or holographic multi-entropy, is geometrically well-defined in the bulk~\cite{Gadde:2022cqi,Gadde:2023zzj,Penington:2022dhr,Iizuka:2025ioc}, but its proposed boundary dual via multi-entropy is subtle: the von Neumann limit $n\to 1$, where $n$ denotes the replica index, is ill-defined in holographic CFTs~\cite{Harper:2024ker} and the bulk replica symmetry is in question~\cite{Gadde:2024taa,Penington:2022dhr}. This motivates an alternative strategy: rather than seeking a direct entropic definition, one may attempt to characterize such quantities operationally --- for instance, via holographic LOCC~\cite{Mori:2024gwe}, and use the associated task to infer their boundary duals.
Specifically, one may ask under which set of operations a quantity of interest is monotone: an answer to this question could hint at the operational significance of the quantity. See Section~\ref{subsec:resource_theory} for a related discussion.

Finally, instead of mapping boundary tasks to bulk quantities, one may invert the perspective and ask what bulk-task feasibility implies for bulk or boundary quantities. 
One manifestation is given by asymptotic quantum tasks and the connected wedge theorem~\cite{May:2019yxi}, which will be discussed in the next problem.

\newcommand{\textTwoTwo}{What are the implications of bulk causality for the information-processing tasks and vice versa?}
\problem{\textTwoTwo}

Holography concerns not only the correspondence between states but also their dynamics. Consider a quantum computer in a bulk subregion located away from the boundary. One may then consider information-processing tasks performed by this device. By holographic duality, the same tasks must, in principle, be implementable solely on the boundary. However, because the bulk-boundary map is nonlocal, a local bulk computation sometimes has to be mapped to a nonlocal computation on the boundary. Reference~\cite{May:2019yxi} sharpened this aspect by formulating a task of information sharing. In the bulk, assuming bulk locality, causality constrains whether two agents can exchange information directly. On the boundary, however, the same task cannot be implemented in a manifestly local manner consistent with these constraints; instead, it is necessarily completed nonlocally. To implement a local computation in a nonlocal manner, one has to make use of preshared correlation as well as classical communication. In holography, such correlations are geometrically captured as a \emph{connected entanglement wedge}, i.e., a geometric connection between two boundary subregions.\footnote{
One should note a potential loophole involving delayed use of preshared correlations in complementary subregions. While Ref.~\cite{May:2019odp} argued that such strategies are ineffective in holographic states due to the absence of GHZ-type entanglement, subsequent work~\cite{Dolev:2022gwj} demonstrated a more subtle protocol based on quantum one-time pads~\cite{Mosca:2000aaw}, where all the required correlations are encoded in complementary subsystems.
The extent to which such constructions evade the connected wedge theorem remains an open question.} Interestingly, the necessity of preshared correlations can be argued purely in quantum information, while in holography it admits a geometric proof via the connectedness of entanglement wedges, which represent a large holographic mutual information~\cite{May:2019odp}. This correspondence is formalized as the \emph{connected wedge theorem}.\footnote{See~\cite{May:2026zdc} for a review accessible to QI readers.} This perspective suggests a bidirectional relation: the achievability of boundary tasks can diagnose bulk causal structure, while bulk causality constrains the feasibility of boundary implementations and hence the required correlation or entanglement structure.

This motivates a more refined question: to what extent can boundary information-processing tasks serve as operational probes of bulk causal structure? The connected wedge theorem relates one information-sharing task to a coarse causal property and to the existence of large boundary correlations. A natural next step is to ask whether analogous tasks can detect finer causal data, such as causal ordering among multiple bulk events, null versus timelike separation, or constraints implied by bulk microcausality. Conversely, one may ask whether such detailed causal data imply sharper limitations on boundary protocols, including no-go results for certain tasks, lower bounds on preshared correlations, or requirements for multipartite entanglement. Establishing such a dictionary would turn bulk causality into a resource-theoretic constraint on boundary information processing.

\newcommand{\textTwoThree}{Can we reformulate QFT correlators in terms of information-theoretic tasks?}
\problem{\textTwoThree}

Task feasibility is often quantified or bounded by entropic quantities such as mutual information. For example, the optimal asymptotic exponent of the type II error in hypothesis testing is given by the relative entropy from the quantum Stein's lemma~\cite{qstine1, qstine2}, while the mutual information characterizes how much local randomness is required to destroy the given correlation~\cite{Groisman:2005dbo}. More generally, protocols such as the fully quantum Slepian-Wolf (FQSW) protocol~\cite{Abeyesinghe:2006xbr,8234697}, often referred to as ``the mother of all protocols," provide an operational meaning of various entropic correlation measures. 
These results are typically formulated in terms of density matrices rather than observables such as correlation functions. 
This raises the question of whether one can establish a direct relation between correlators and operational quantities or tasks. 	
In particular, can one identify monotones that quantitatively relate correlation functions to task feasibility, for instance, through success probabilities or error exponents? 

Correlation functions are fundamental ingredients in QFT. In particular, they are related to scattering amplitudes through the Lehmann-Symanzik-Zimmermann (LSZ) formula. 
In QG, correlation functions also play a key role; for example, in celestial holography, they connect observables on the celestial sphere to scattering amplitudes in asymptotically flat spacetime~\cite{Pasterski:2016qvg,Raclariu:2021zjz,Pasterski:2021raf}.
Time-ordered correlators, in particular, encode information about scattering singularities.  However, in holographic settings, bulk singularities do not always coincide with singularities in boundary correlators~\cite{Maldacena:2015iua,Komatsu:2020sag,Caron-Huot:2025hmk}. It remains an open problem how to actually `see’ the bulk singularities from the boundary correlation functions. On the other hand, results such as the connected wedge theorem suggest a relation between bulk scattering processes and the boundary mutual information. This provides evidence that correlation functions may be linked to entropic quantities, and hence to operational tasks.

In the context of scrambling, out-of-time-ordered correlators (OTOCs)~\cite{1969JETP...28.1200L,Maldacena:2015waa,Shenker:2013pqa,Roberts:2014isa,Hosur:2015ylk,Roberts:2016hpo,Garcia-Mata:2022voo}
play a central role in characterizing the chaotic nature of its local interactions. They are defined as
\begin{equation}
C(t) = - \ev{\comm{W(t)}{V(0)}^2} \quad\text{or}\quad F(t)=\ev{V(0)W(t)V(0)W(t)}
\end{equation}
for two operators $V, W$ with a suitable Euclidean regularization. At sufficiently late times, the time-ordered contributions to $C(t)$ factorize, and the remaining non-factorizing part is captured by $F(t)$ (and its counterpart with $V$ and $W$ exchanged), which typically oscillates around a constant value.
Before saturation, OTOCs typically exhibit exponential growth, $C(t)\sim e^{\lambda_L t}$, characterized by the Lyapunov exponent $\lambda_L$. While this description is valid in a semiclassical regime, Krylov complexity could serve as a probe beyond this regime~\cite{Parker:2018yvk}. Both OTOCs and Krylov complexity are defined in terms of correlation functions and are probes of operator growth. In particular, Krylov complexity admits an interpretation in terms of Nielsen/circuit complexity~\cite{Craps:2025kub}. Nevertheless, their operational meaning through information processing tasks remains largely unclear.

The timescale at which OTOCs saturate, $t\sim \lambda_L^{-1}\log S$, is known as the scrambling time.\footnote{Here $S$ denotes the entropy, which scales as $N^2\sim G_N^{-1}$ in holography. In terms of $F(t)\sim 1 -\frac{1}{N^2}e^{\lambda_L t}$, the exponential growth arises from the subleading correction.} 
In the context of the black hole information paradox, particularly in the Hayden--Preskill model~\cite{Hayden:2007cs}, the Petz-type recovery map becomes approximately effective after the scrambling time. One manifestation is given by the Yoshida--Kitaev decoder~\cite{Yoshida:2017non}. 
In holography, this corresponds to the entanglement wedge reconstruction of the region including the island region, an emergent bulk subregion in the black hole interior. This suggests a possible direct connection between scrambling diagnostics and operational tasks such as information recovery. 
Indeed, recovery/distillation protocols sometimes admit a gravitational interpretation as traversable wormholes. It has been recently argued that the time derivative of OTOCs lower bounds the quantum channel capacity of the Gao-Jafferis-Wall traversable wormhole, assuming coherent information approximates the capacity well~\cite{Lu:2026dns}. This result offers a concrete step toward relating correlators such as OTOCs to operational quantities and information-processing tasks.

\newcommand{\textTwoFour}{Can we identify any task wherein QG effects play an important role?}
\problem{\textTwoFour}

So far, the discussion has largely focused on the semiclassical regime. However, since AdS/CFT is expected to hold beyond the infinite-$N$ limit, where $N$ denotes the large-$N$ parameter controlling the number of boundary degrees of freedom,  it is natural to ask whether there exist tasks whose performance is enhanced or degraded by genuine QG effects. 
At finite $N$, the (effective) dimension of the Hilbert space is expected to be finite in contrast to the infinite-$N$ semiclassical limit, where the entropy is unbounded.
Since many information-processing tasks utilize entanglement as a resource, this change may lead to a large gap in achievable performance, for instance, in entanglement-assisted channel capacities over noisy quantum channels~\cite{Bennett:1999hf,Bennett:2002owy,Devetak:2005lvq,Hsieh:2010smw}. At the same time, it is not clear that all entanglement present in the state is operationally accessible as a resource. A systematic understanding of these operational aspects is still in its infancy. 
It is therefore desirable to formulate these questions in a more abstract algebraic framework, for example, in terms of operator-algebra quantum error-correcting codes (OAQECC).
We refer to Sections~\ref{sec:QEC} and~\ref{sec:infinite_dim} for related discussions.

Apart from holography, gravity-mediated (or gravity-induced) entanglement (GME/GIE), as proposed in the Bose-Marletto-Vedral (BMV) setup~\cite{Bose:2017nin,Marletto:2017kzi}, provides a potential low-energy probe of the quantumness of gravity.\footnote{See~\cite{Marletto:2024ltk,Sahdo:2025cyr} for reviews of GME and~\cite{Marletto:2017upo} for an introduction to non-experts. Numerous works discuss experimental implementations; see references therein.} Here, the goal is to probe the non-classicality of gravity from an entanglement generation between two massive objects. 
This is precisely where operational tasks become essential: they are needed to verify what the resulting entanglement actually signifies.
This has provoked discussions on whether such entanglement necessarily implies a quantum mediator, or whether classical or hybrid theories can account for the phenomenon~\cite{Hall:2017nzl,Marshman:2019sne,Aziz:2025ypo,Marletto:2025fpm,Marletto:2025asw,Mitrakos:2026xlr,DiBiagio:2025twt,Christodoulou:2018cmk}. 
Addressing this issue requires a broader, careful operational formulation, for instance in terms of general witness theorems~\cite{Marletto:2017pjr,Marletto:2017wuj} or C$^\ast$-/von Neumann algebras~\cite{Ludescher:2025kko}.

\subsection{Can we apply resource theory to holography and QFT?}
\label{subsec:resource_theory}

Operational questions are naturally framed in the language of \emph{resource theories}~\cite{Chitambar:2018rnj,Lostaglio:2019ruq,Coecke:2014svf,kuroiwa2020generalquantum}.
A resource theory is a theoretical framework specified by a set of \emph{free operations} $\mathcal{O}$ (those that can be implemented at no cost); typically free operations determine a set of \emph{free states} $\mathcal F$ (those that are easy to prepare) by requiring that every free operation maps $\mathcal F$ into itself.
One then asks which state transformations, simulations, or operational tasks are achievable using $\mathcal{O}$, possibly supplemented by non-free resource states.
A standard example is entanglement theory: the free operations are LOCC, the free states are separable states, and entanglement serves as a resource enabling nonlocal tasks.
Operational quantities in this setting correspond to conversion rates (e.g., distillable entanglement) or resource costs (e.g., entanglement cost).
In nonlocal games, one may ask for the maximum winning probability achievable with a given amount of entanglement under local operations assisted by shared randomness, or conversely, for the minimum entanglement required to achieve near-perfect success~\cite{Slofstra:2018crq,Cleve:2010xmq}.

\newcommand{\textTwoFive}{How should we choose free states/operations in the  application of resource theories to QFTs?}
\problem{\textTwoFive}

In QFT, the choice of $\mathcal{O}$ is subtler, as one must account for locality, causality, and, in some cases, spacetime symmetries.
One may restrict attention to operations localized within spacetime regions (or, precisely, to von Neumann algebras associated with such regions), to allow classical communication only along causal curves, and to impose additional constraints such as energy, charge, or finite-time implementability.
These choices are not merely auxiliary; rather, they capture the operational meaning of ``access to a subregion'' or the ``limitations of a physical observer''~\cite{Bouland:2019pvu,Aaronson:2016vto,Brown:2019rox,Engelhardt:2021mue,Engelhardt:2024hpe,Yang:2023zic,Engelhardt:2024lnd,Leone:2025qkj,Gu:2023qqq,Meyer:2025dmr}. 
An illustrative example of how operational constraints shape physics is the firewall near a black hole horizon, which involves a large blueshift under boosts and can be viewed as a complexity cutoff on the allowed operations~\cite{Harlow:2013tf}.

Alternatively, one may consider a more abstract notion of $\mathcal{O}$ defined in theory space, for instance, in terms of couplings in a Lagrangian or Hamiltonian.
A canonical example is coarse-graining via renormalization-group (RG) transformations, which integrate out short-distance degrees of freedom and yield effective descriptions at longer length scales.
Although the notions of free operations and free states are less clear in this setting, viewing RG flow through the lens of resource theory may offer a systematic and rigorous perspective on renormalization, complementary to approaches such as exact or functional RG~\cite{Bagnuls:2000ae,Gies:2006wv,Costello:2007ei,Buchholz:1995gr}. 
We will consider this problem in more detail in Problem \problink{2}{7}.

\newcommand{\textTwoSix}{Can we use resource theory to find new physical monotones and operations?}
\problem{\textTwoSix}

Given physically motivated operation classes in QFT (causal operations, operations with a compact support in spacetime, thermal operations, energy-bounded operations, symmetry-covariant maps, etc.), what are the associated monotones and complete families of constraints (i.e., necessary and sufficient conditions) governing state transformations?
Furthermore, for quantities $\mathcal{Q}$ appearing in holography, such as reflected entropy~\cite{Dutta:2019gen}, Markov gap~\cite{Hayden:2021gno,Zou:2020bly}, or multi-entropy~\cite{Gadde:2022cqi,Gadde:2023zzj,Penington:2022dhr}, can one identify a physically well-motivated class of free operations $\mathcal{O}$ for which $\mathcal{Q}$ serves as a resource monotone? More concretely, can $\mathcal{O}$ be derived from fundamental QFT principles (locality, covariance, unitarity, energy constraints) in such a way that $\mathcal{Q}$ acquires a clear operational meaning? 
A further question concerns the uniqueness and robustness of such choices of $\mathcal{O}$.
Is there a natural set of axioms characterizing ``physically admissible'' free operations, and can one establish uniqueness or universality results for the choice of $\mathcal{O}$ associated with a given quantity $\mathcal{Q}$?

\newcommand{\textTwoSeven}{Can we formulate a resource theory of renormalization group flow?}
\problem{\textTwoSeven}

A natural direction is to reinterpret \emph{renormalization group flow}, a fundamental concept at the heart of QFT and statistical mechanics, within a resource-theoretic framework.
In standard resource theories, free operations are typically described by completely positive trace-preserving (CPTP) maps, and free states form a convex set.
In contrast, in the present context, RG flows play the role of free operations, while fixed-point theories correspond to free states.
This departure from the standard framework already raises conceptual challenges, particularly in identifying appropriate monotones.

Monotones play a central role in any resource theory. The usual strategy is to specify free operations and then characterize functions that are monotonic under them. A classic example of an RG monotone in QFT is the $c$-function in two dimensions~\cite{Zamolodchikov:1986gt}, which decreases monotonically along the RG flow, reflecting irreversibility.
Higher-dimensional analogues include the $a$-theorem in four dimensions~\cite{Komargodski:2011vj}.
More generally, certain higher-derivative couplings in the effective field theories exhibit monotonic behavior under RG flow~\cite{Arkani-Hamed:2021ajd,Liao:2025npz}.
From this perspective, QFT admits analogues of resource theories in which RG flow plays the role of the allowed operations, while quantities such as the $c$- and $a$-functions play the role of monotones.
A notable feature of RG monotones is that the monotonicities typically rely on physical assumptions such as unitarity and Lorentz invariance.
Although some RG monotones admit an information-theoretic interpretation in terms of entanglement entropy~\cite{Casini:2006es,Casini:2012ei,Casini:2017vbe,Lashkari:2017rcl,Nishioka:2018khk}, a unified structural understanding remains incomplete.
This naturally raises the question of how such physical constraints should be incorporated into a resource-theoretic formulation.

This perspective extends naturally to holography.
In AdS/CFT, the emergent radial direction is associated with the energy scale of the boundary theory, a relation often referred to as holographic RG~\cite{Bianchi:2001kw,deBoer:1999tgo,Heemskerk:2010hk,Kiritsis:2014kua,Fukuma:2002sb,Kim:2025mrp,Skenderis:2002wp}.
This suggests interpreting geometric quantities, such as the area of quantum extremal surfaces (i.e., holographic entanglement entropy), as candidates for the gravity dual of resource monotones.
A corresponding notion of free operations is less clear. 
One possible candidate, which may belong to the broader class of resource non-generating operations, is coarse-graining of the boundary theory by increasing the UV cutoff~\cite{McGough:2016lol,Hartman:2018tkw}.
Making these ideas precise in a holographic setting remains an open problem.
This naturally leads to a broader question:

\newcommand{\textTwoEight}{Can we construct a holographic resource theory? Namely, what is a gravity dual of boundary resource theory, in which free operations and monotones admit geometric or gravitational descriptions?}
\problem{\textTwoEight}

In holography, a natural candidate for free states $\mathcal{F}$ is given by holographic states, i.e., quantum states admitting semiclassical gravity duals. 
Free operations $\mathcal{O}$ should then map such states to themselves; one possible class is given by semiclassicality-preserving maps. 
While identifying such operations directly in the boundary theory is challenging, in the bulk effective description one may consider geometric operations such as small excitations or shockwave perturbations. On the boundary, these are expected to correspond to single-trace operators or suitably constrained heavy operators.

Focusing on a fixed semiclassical background, one may take free operations to be those that preserve the code subspace, in the sense of holographic quantum error correcting codes~\cite{Pastawski:2015qua,Almheiri:2014lwa}. From this viewpoint, the problem could be recast in terms of quantum resource correction~\cite{Byrd:2025nms}, providing a potential bridge between holography and resource-theoretic frameworks.

\section{Observer}\label{sec:observer}
Observers play a subtle but central role throughout quantum theory. The concept of an observer enters the theory through at least three related but distinct roles, each highlighting a different physical requirement:

\begin{itemize}
\item \textbf{(Observer as a reference frame)} The first role is that of a clock or, more generally, a reference frame relative to which time evolution, the spatial localization of physical systems and observables, and physical change are defined. In standard QI, this structure is usually fixed in advance, as in a quantum circuit with a prescribed ordering of gates. In relativity and quantum gravity, by contrast, temporal order and localization may be frame-dependent or relational, and in generally covariant systems, there may be no preferred external time at all~\cite{Page:1983uc,Rovelli:1990ph,Giacomini:2017zju}.

\item \textbf{(Observer as an operational limitation)} The second role is a limitation on accessible operations and information. An observer may be characterized by restricted measurements \cite{Bartlett:2006tzx,Fewster:2024pur}, finite resolution \cite{Wehrl:1978zz,Safranek:2019nwk,Safranek:2020tgg}
, computational complexity \cite{Harlow:2013tf,Susskind:2014rva,Bouland:2019pvu,Kim:2020cds}, causal accessibility \cite{Hubeny:2012wa,Czech:2012bh,Dong:2016eik}, or an information cutoff~\cite {Page:1993wv,Hayden:2007cs,Engelhardt:2017aux}, all of which affect which observables and entropic quantities have operational meaning.

\item \textbf{(Observer as a detector)} The third role is a detector or measurement apparatus. This is the familiar operational role of the observer in quantum mechanics, but it also has concrete realizations in QFT, for example, through localized detector models and through the asymptotic observers implicit in scattering theory~\cite{Unruh:1976db,Haag:1958vt}.
\end{itemize}

These roles provide a common language for connecting QI, QFT, and QG. QI emphasizes agents, circuits, and reference frames, as well as causal order and operational restrictions on tasks; relativistic quantum information treats spacetime localization as an explicit resource, as exemplified in detector-based protocols or position-based cryptography~\cite{Peres:2002wx,Buhrman_2014}. QFT incorporates observers through measurement models, local algebras of causal diamonds, and asymptotic scattering data, while also raising subtle questions about which observables are physically accessible beyond perturbation theory.

In this section, we examine the role of observers in QI and QG. We first address the challenge of articulating the concept of observers in QG (Section~\ref{subsec:obs-concept}), and then turn to the question of how to incorporate the dynamics of observers (Section~\ref{subsec:obs-dynamics}).

\subsection{How should we articulate the concept of observers in QG?}\label{subsec:obs-concept}

The role of observers becomes subtler and more prominent in the presence of gravity. In QG, background independence removes the possibility of an external observer with a preferred clock or frame, and gauge invariance often requires observables to be relational or gravitationally dressed. The recent discussions of observers in holography and closed universes should be understood as a modern manifestation of a long-standing problem: how the physical content of a quantum theory is contingent upon the clocks, detectors, reference frames, and operational limitations through which it is accessed.

We organize this discussion around the following three open questions.

\newcommand{\textThreeOne}{
What reference structures are required to define observables in quantum gravity?}
\problem{\textThreeOne}

In non-gravitational QFT, observables are usually specified intrinsically in terms of local operators and their correlation functions. Although the choice of a laboratory frame may be important operationally, it is not normally part of the definition of the theory. In a gravitational theory, this separation becomes much less clear. Since diffeomorphisms act as gauge redundancies, a local quantity defined at a coordinate point is not by itself a gauge-invariant observable. To define a physical observable, one must either fix a gauge or define the observable relationally with respect to physical reference data, such as an asymptotic frame, a clock, a detector, a worldline, or other degrees of freedom that can serve as a reference system.

In this sense, an observer enters quantum gravity as the physical or operational structure relative to which localization, time evolution, and measurement are defined. A reference frame specifies not only coordinates, but also an algebra of accessible observables and, in many situations, a reference state in which a semiclassical description is valid.
The unresolved issue is not simply how to construct more examples of dressed or relational observables. Rather, it is to identify the minimal data that must be specified in order for an observable algebra to have a semiclassical geometric interpretation. In many examples, this data includes not only a reference frame in the ordinary spacetime sense, but also an accessible algebra, a reference state, and a code subspace on which the reconstruction is valid. Thus, the problem is to understand whether there is a general principle that assigns an effective algebra of bulk observables to each admissible reference structure, and when two observer-centered descriptions based on different reference structures should be regarded as equivalent descriptions of the same physics.

This issue is already visible in AdS/CFT. The boundary CFT gives a complete microscopic description, but bulk observables are not automatically defined as universal operators acting on the entire CFT Hilbert space. Rather, a bulk reconstruction is usually tied to a choice of reference structure: an asymptotic frame, an accessible algebra, and often a reference state. For example, operators in the exterior of an anti-de Sitter (AdS) black hole are naturally associated with the algebra accessible to an asymptotic observer, while operators behind the horizon require additional input. In the Papadodimas--Raju construction, for example, interior mirror operators are defined relative to a suitable equilibrium state and a small algebra~\cite{Papadodimas:2013jku,Papadodimas:2013wnh}. Thus, the interior operator is not a universal state-independent CFT operator on the full Hilbert space, but rather an effective observable defined within a particular reference frame.

The same point appears in a more geometric form in the relation between entropy and area. The Ryu--Takayanagi formula relates boundary entanglement entropy to the area of a bulk extremal surface~\cite{Ryu:2006bv,Ryu:2006ef,Hubeny:2007xt}. However, entanglement entropy is a nonlinear functional of the boundary state, and the relevant extremal surface can itself depend on the state or density matrix. Therefore, the area should not be understood as a single state-independent linear operator on the full microscopic Hilbert space. It is better viewed as an effective geometric observable defined within a semiclassical reference state~\cite{Almheiri:2016blp}. This again shows that geometric observables require specifying the class of states and the reconstruction scheme in which they are meant to act.

A similar lesson is suggested by the double-scaled Sachdev–Ye–Kitaev (DSSYK) model, particularly by recent analyses of its finite-$N$ bulk Hilbert space \cite{Lin:2022rbf,Miyaji:2025dssyk}. In the strict large-$N$ double-scaling limit, the model admits an emergent geometric description in which certain basis states and operators can be interpreted as describing the bulk geometry. At finite $N$, however, the microscopic Hilbert space is finite-dimensional, while the naive geometric construction, if extrapolated to finite $N$, produces infinitely many formal bulk states. Therefore, these geometric states cannot all be independent. Instead, there are linear relations and null states among them, reflecting the finite-dimensional nature of the exact Hilbert space. As a result, the bulk geometric observable should not be regarded as a single state-independent operator acting on the entire microscopic Hilbert space. It is better understood as an effective observable defined only within a suitable semiclassical regime, for example near a chosen reference state. In this way, finite-$N$ DSSYK provides a concrete toy model for the general lesson that semiclassical geometric observables acquire a precise meaning only after specifying the reference structure in which they are represented.
Therefore, the question of observables in QG is not only which operators exist, but also what reference structure makes those operators well-defined.

The issue becomes even sharper in de Sitter space, where there is no spatial boundary with a standard Hilbert-space interpretation. Different observers, such as those associated with different static patches, have access to different sets of observables, and there is no obvious global algebra that simultaneously captures all such perspectives. In the static patch, observables can be gravitationally dressed to the worldline of an observer, and the resulting algebra has been argued to be a Type II$_1$ von Neumann algebra whose entropy reproduces generalized entropy up to a state-independent additive constant~\cite{Chandrasekaran:2022cip,Witten:2023qsv}. In this setting, the observer is not merely an interpretational device, but part of the definition of the observable algebra.

A closely related perspective is provided by quantum reference frames. In this approach, a reference frame is not treated as an external classical structure, but as a quantum system whose degrees of freedom describe relational observables. Changing the reference frame corresponds to a transformation between different relational descriptions of the same underlying system~\cite{Aharonov:1984zz,Rovelli:1990pi,Bartlett:2006tzx,Giacomini:2017zju,Vanrietvelde:2018dit,Hoehn:2021,DeVuyst:2024khu,DeVuyst:2024fxc,Hoehn:2019fsy}. From this viewpoint, what appears as an observer may arise from an internal choice of relational degrees of freedom, rather than from an external structure added to the theory.

Thus, the central problem is not merely to construct more bulk operators. It is to specify the reference structure relative to which such operators are meaningful. In this sense, the problem of observables in quantum gravity is inseparable from the problem of reference frames. This naturally leads to the next question: whether the observer or reference structure is a fundamental input, emergent effective data, or a redundancy of a relational description.

\newcommand{\textThreeTwo}{Is an observer an emergent concept? Is an observer an additional structure required for consistency, or merely an artifact of the description?}
\problem{\textThreeTwo}

There are several possible answers (which are by no means mutually exclusive). One possibility is that an observer is an emergent concept. On this view, observers appear only in a semiclassical regime where geometry, causal patches, horizons, and localized detectors are meaningful. The observer would then not be part of the fundamental formulation, but an effective concept that arises in a particular limit.

A second possibility is that an observer represents additional structure required for consistency. If gauge-invariant observables in gravity can be defined only after specifying a reference system---for example, a worldline, an asymptotic frame, a causal patch, an accessible algebra, or a reference state---then the observer is not optional. It is part of the data required to extract physical predictions from the theory.

A third possibility is that an observer is an artifact of a particular description. In approaches based on quantum reference frames, one attempts to describe physics in a fully relational manner, where no single observer frame is fundamental, and different observer-centered descriptions are related by transformations within a single underlying framework~\cite{Aharonov:1984zz,Rovelli:1990pi,Bartlett:2006tzx}. From this perspective, observer dependence may reflect a redundancy in the description rather than an additional physical ingredient.

Black hole interiors suggest that the correct answer may involve elements of all three. From the boundary point of view, the microscopic Hilbert space is perfectly well-defined. However, the semiclassical interior is not described by a universal, state-independent operator algebra on the full finite-$N$ Hilbert space. Instead, it is reconstructed only within a restricted domain: a small algebra, a reference state, and a class of probes for which effective field theory is valid. 
This is closely related to non-isometric reconstruction. When the effective Hilbert space of semiclassical bulk effective field theory (EFT) is larger than the fundamental Hilbert space that encodes it, some effective states are null or become indistinguishable under the allowed class of simple operations~\cite{Akers:2022qdl}. Interior reconstruction must then be tied to the restricted set of states and operations accessible to a given observer.

Closed universes sharpen this point further. In a gravitational path integral without an asymptotic boundary, the putative nonperturbative Hilbert space can be extremely small, and in some formulations effectively one-dimensional. At the same time, the experience of an internal observer may be described by an emergent quantum-mechanical Hilbert space whose size is controlled by the observer's own degrees of freedom~\cite{Harlow:2025qmo}. Related discussions connect this question to holographic maps, observer complementarity, Hartle--Hawking states, and baby-universe sectors~\cite{Abdalla:2025gzn,Akers:2025obs,Nomura:2025closed,Engelhardt:2025azi,Harlow:2026alpha,Zhao:2026itfrombit,Higginbotham:2025dvf}.

Our working viewpoint is that these possibilities are different aspects of the same phenomenon. An observer need not be a new fundamental degree of freedom added to the theory. At the same time, it is not merely a linguistic artifact. Rather, an observer is an effective relational structure that selects the algebra, state, and operations in which a semiclassical geometry is represented. The remaining problem is to understand how such structures emerge from the microscopic theory, what their domain of validity is, and when two different observer-centered reconstructions describe the same underlying quantum state.

This leads to a further question. If the observer is an effective or emergent structure, then it cannot be assumed to have unlimited access to the microscopic state. Its finite resolution, finite lifetime, causal patch, and computational power should determine which parts of the bulk geometry are operationally meaningful. This motivates the next problem.

\newcommand{\textThreeThree}{Can we quantify fundamental limits to what an observer can learn about the bulk geometry?}
\problem{\textThreeThree}
In Problems \problink{3}{1} and \problink{3}{2}, we argued that defining observables in quantum gravity requires specifying a reference structure: a reference frame, an observer algebra, a reference state, or a code subspace. Once such a structure is specified, the next question is not only what information is present in the microscopic state, but what information can be extracted by that observer. In particular, information may exist in principle without being operationally accessible to an observer with finite computational or experimental resources. Thus, the observer's capabilities help determine which bulk regions or geometric degrees of freedom are physically accessible.
This gives an operational way to formulate the problem of emergent geometry. Rather than asking only whether a semiclassical region is encoded in the microscopic theory, one should ask which class of observers can reconstruct it, by which operations, and within which code subspace. In this sense, computational complexity, non-isometric reconstruction, pseudorandomness, and finite-time detector observables are not merely technical complications. They are part of the definition of the effective bulk geometry seen by a bounded observer.

This question is relevant for an observer with finite computational or operational resources. Here we focus on complexity-theoretic limitations.\footnote{More radical limitations, such as undecidability or non-computability of certain questions in quantum gravity, would represent a different class of obstructions and will not be discussed here. However, some undecidable problems are related to Tsirelson's problem, explained briefly in Problem \problink{5}{1}~\cite{Ji:2020apg,Perales-Eceiza:2024qhd}.} A sharp example is the Harlow--Hayden decoding task for an old black hole. Consider a late Hawking mode $B$ emitted after the Page time and the early radiation $R$. Assuming unitary black hole evaporation, the microscopic state is expected to contain, within $R$, a subsystem that purifies $B$, up to the usual semiclassical and typicality assumptions. The operational question is whether an exterior observer, with coherent quantum control over $R$ and with knowledge of the microscopic encoding or evaporation dynamics, can implement a decoding map on $R$ that extracts a small subsystem $R_B$ entangled with $B$. This is a single-copy recovery problem, not an entanglement-distillation protocol in the usual many-copy LOCC sense. The Harlow--Hayden argument suggests that such a decoding operation requires a time exponential in the black hole entropy~\cite{Harlow:2013tf}. Thus, for a computationally bounded observer, the interior partner mode may be present in principle in the microscopic state but remain operationally inaccessible~\cite{Brown:2019rox,Engelhardt:2021qjs,Yang:2023zic,Engelhardt:2024hpe,Akers:2022qdl}.

There is also a related limitation associated with the reconstruction of operators behind the horizon.
In state-dependent approaches to the interior, the boundary representative of an interior mode is defined only after choosing a reference state, or a sufficiently small code subspace, as in the construction of mirror operators~\cite{Papadodimas:2013wnh,Papadodimas:2013jku}.
Thus, even if the microscopic state contains enough information to describe a smooth interior, there need not be a single state-independent boundary operator that extracts the same interior observable on all semiclassical states.
This is natural from the perspective of non-isometric codes: when the effective interior Hilbert space is larger than the fundamental one, interior EFT can be valid only within a restricted set of states and probes~\cite{Akers:2022qdl}.
This is the algebraic version of a tension familiar from the firewall and state-dependence discussions.
A smooth horizon requires interior partner operators for exterior modes, while realizing such operators as state-independent CFT operators on the full finite-$N$ Hilbert space is highly constrained~\cite{Almheiri:2012rt,Marolf:2013dba,Harlow:2014yoa}.
Therefore, the limitation on learning the interior is not only computational; it also concerns the range of states on which a given reconstruction map is valid.

A complementary obstruction arises from computational indistinguishability.
Assuming quantum-secure pseudorandom functions and pseudorandom permutations, as instantiated for example from quantum-secure one-way functions, there exist ensembles of efficiently preparable states have parametrically small entanglement but are computationally indistinguishable, for polynomial-time quantum observers with polynomially many copies, from highly entangled Haar random states.
Such spoofed entanglement is called pseudoentanglement~\cite{Aaronson:2022flk}, and possible bulk duals may be constructed from pseudorandom states~\cite{Engelhardt:2024lnd}.

In some cases, even with an exponential power, the entanglement may not be spoofed. For example, in~\cite{Antonini:2023hdh}, a closed universe is entangled with two asymptotic boundaries. In~\cite{Mori:2025jej}, using a Haar random encoding model, the baby universe emerges as a result of the failure of complementary recovery, indicating an observer residing in one of the asymptotic boundaries can never access the baby universe's logical degrees of freedom even with the exponential complexity. 

Ultimately, the computational and operational power of the observer is crucial for understanding the physical phenomena described by emergent semiclassical geometries and algebras.\footnote{See~\cite{Kudler-Flam:2025cki,Liu:2025cml,Liu:2025ikq,Engelhardt:2025azi} for some recent approaches in this direction.}

\subsection{Can observers be explicitly incorporated into the dynamical descriptions of spacetime?}\label{subsec:obs-dynamics}
\newcommand{\TY}[1]{\textcolor{red}{#1}}
The preceding discussion treats observers primarily as reference structures that determine what counts as an observable. However, in any operational setting, an observer is also a physical system that interacts with quantum fields through localized measurements. This suggests a second, more dynamical question: how should observables be defined when the clock, detector, and information-recording apparatus are themselves part of the gravitating system? This question becomes especially sharp in black hole evaporation, where the relevant data are not only asymptotic in- and out-states, but also the time-resolved signal registered by finite-size detectors.

In the standard formulations, an observer is treated as an external probe to the theory (see e.g.\ \cite{Geng:2025bcb,Speranza:2025joj,Chen:2024rpx} for exceptions). However, an observer in QG cannot be completely external, and we need to take its own dynamics into account. Such dynamical aspects of observers are often not treated systematically in the literature.

There are two related aspects of this problem. 
First, the observer carries energy and information, and therefore backreacts on the geometry and on the entanglement structure. 
Second, the observables accessible to a finite-lifetime observer are naturally finite-time detector observables rather than idealized asymptotic $S$-matrix elements. 
In black hole evaporation, these two issues meet: the physically relevant question is not only what the final asymptotic state is, but what time-resolved signal a detector records while the black hole is evaporating.

\newcommand{\textThreeFour}{If the observer is a dynamical degree of freedom within the theory, how does the measurement process modify the semiclassical geometry and the observable algebra?
}
\problem{\textThreeFour}

In QG, we must treat the observer not merely as a probe but as a physical object equipped with a non-trivial stress-energy tensor, thus backreacting on the spacetime. 

For example, in the context of the Island formula \cite{Penington:2019npb,Almheiri:2019psf,Almheiri:2019hni,Almheiri:2019qdq,Penington:2019kki2,Goto:2020wnk}, the location of the quantum extremal surface (QES) depends on the bulk entanglement entropy. If the measurement is modeled as an interaction with a dynamical apparatus, the combined state of the apparatus and the bulk fields can change the bulk entropy term entering the generalized entropy. 
This is closely related to evaporating black hole models in which an AdS gravitational region is coupled to a nongravitating bath or to an asymptotically flat region that collects the radiation. 
In these setups, the entropy of the radiation is computed in an enlarged system, and the resulting generalized entropy can develop nontrivial QES saddles and replica wormholes~\cite{Almheiri:2019psf,Goto:2020wnk}. 
Likewise, when a measuring apparatus is included as part of the quantum system, the location of the QES, and hence the entanglement wedge, may become sensitive to how the apparatus and its records are included, traced out, or conditioned upon. In such situations, the location of the QES, and hence the entanglement wedge, may become sensitive to the measurement setup. 
This should not be understood as an instantaneous collapse of the geometry, but rather as a statement about how the operationally accessible algebra and its semiclassical reconstruction depend on the inclusion of the measuring apparatus. 
In the replica path-integral description of QES, the same dependence can appear as a change in which degrees of freedom are replicated, traced over, or conditioned upon, and hence as a change in the saddle-point competition between disconnected saddles and replica-wormhole saddles~\cite{Qi:2021sxb}. 
Related issues also arise in discussions of black hole interiors and the black hole final-state projections, where the reconstruction of the semiclassical interior can depend on the state, on the measurement history, and on how an observer probes the geometry~\cite{Papadodimas:2015xma,Almheiri:2025mwp,Yoshida:2022srg}.

\newcommand{\textThreeFive}{Can we formulate a measurement theory in the presence of stringy interactions?
What consequences does it have?}
\problem{\textThreeFive}

In this problem, we mainly focus on the ``observer as a detector'' mentioned at the beginning of this section, while also discussing the ``observer as an operational limitation.''

When formulating a theory of measurement in quantum theory, one introduces operators corresponding to observables and detectors, and examines their commutation relations.
For example, when discussing the causality of signal transmission in relativistic quantum theory, one typically adopts the Unruh-DeWitt type model \cite{Unruh:1976db,DeWitt:1980hx} and introduces an interaction Hamiltonian between detectors and quantum fields:
\begin{align}
    \hat{H}_{\text{int.}}\sim \hat{D}(t)\otimes\hat{\phi}(t,\mathbf{x}).
\end{align}
The microscopic causality of the field appearing here is expressed by the vanishing of the commutator
\begin{align}
    \expval*{[\hat{\phi}(t,\mathbf{x}),\hat{\phi}(t',\mathbf{x}')]},
\end{align}
at spacelike-separated points.
The proper formulation of measurement in relativistic field theory, and the consequent avoidance of superluminal propagation, have been actively discussed in various contexts
\cite{sorkin:1993gg,Beckman:2001qs,Louko:2006zv,Satz:2006kb,
Fewster:2018qbm,Borsten:2019cpc,Bostelmann:2020unl,Polo-Gomez:2021irs,deRamon:2021nry,deRamon:2023qcp}.

We now consider a natural extension of this framework.
Can such a theory of measurement be extended to cases in which stringy interactions are present?
One might expect this to be achieved by replacing the detector and the local operators appearing in field theory with string fields, introducing an interaction of the following form,
\begin{align}
    \hat{\Psi}[X^\mu]\otimes\hat{\Phi}[X^\nu] \quad (?)
\end{align}
and examining the corresponding commutation relations (cf. \cite{Hata:1996hd}),
\begin{align}
    \expval{[\hat{\Phi}[X^\mu],\hat{\Phi}[X^\nu]]} \quad (?)
\end{align}
However, the properties of point particles and strings are fundamentally different.
While a point particle is described by a local field at a spacetime point $x^\mu$, a string is described by a functional of the embedding coordinates $X^\mu(\sigma)$ along the string.
In more physical terms, this means that the finite size of a string imposes an intrinsic limitation on temporal and spatial resolution.
If one tries to construct a theory of measurement in the presence of such stringy interactions, the intrinsic resolution limit would modify the conventional theoretical framework.
What would such a new theoretical framework look like, and what kinds of insights or consequences would it yield?
These questions will likely require an integration of theories of measurement, information, and quantum gravity.

How should we confront this problem?
One possible approach is to employ stringy toy models.
Historically, various models have been proposed to incorporate the finite size of strings.
For example, the generalized uncertainty principle
\cite{Amati:1987wq,Amati:1988tn,Garay:1994en,Kempf:1994su}:
\begin{align}
    [\hat{x},\hat{p}] = i\hbar(1+\ell^2\hat{p}^2),
\end{align}
the non-commutative theories
\cite{Witten:1985cc,Seiberg:1999vs}:
\begin{align}
    [\hat{x}^\mu,\hat{x}^\nu] = i\ell^2\theta^{\mu\nu},
\end{align}
and the spacetime uncertainty principle
\cite{Yoneya:1987gb,Yoneya:1989ai,Yoneya:1997gs,Yoneya:2000bt,PhysRevLett.78.1219}:
\begin{align}
    \Delta X\Delta T \gtrsim \ell^2.
\end{align}
These models incorporate an extra uncertainty arising from the finite size of strings, in addition to the ordinary uncertainty of quantum theory.
A natural first attempt is to construct a theory of measurement in such quantum-mechanical models.

One possible application is to measure the arrival time of string wave packets in string scattering. If the measuring processes are corrected by string effects, the measured value is expected to be affected by the extra uncertainty, which may lead to different results from those in ordinary field theories.

The resulting insights may offer a new perspective on black hole physics.
This is because, based on the black hole-string correspondence \cite{Horowitz:1996nw,Horowitz:1997jc}, the formation and evaporation process of black holes are expected to be described as string scattering processes \cite{Amati:1999fv,Veneziano:2004er,Amati:2007ak,Addazi:2016ksu}.
In this picture, the arrival time of a string wave packet would correspond to the emission time of Hawking radiation.
Whether this emission time contains stringy corrections remains an open problem.
This question is motivated by previous studies suggesting that stringy corrections appear more significantly in the emission time rather than in the radiation spectrum of Hawking radiation, leading to the termination of Hawking radiation \cite{Ho:2022gpg,Akhmedov:2023gqf,Chau:2023zxb,Ho:2023tdq,Ho:2026xhu,Cheng:2026rsm}.
A more complete understanding of the arrival time of string wave packets, grounded in a theory of string measurement, may make it possible to derive the termination of Hawking radiation more rigorously from string scattering amplitudes.

\section{Quantum error correction}\label{sec:QEC}

The profound connection between quantum error correction (QEC) and QG has emerged as a cornerstone of holography. In this framework, the emergence of bulk spacetime geometry from boundary conformal field theory is understood as a realization of a quantum error-correcting code: local bulk operators are encoded into the boundary Hilbert space in a highly nonlocal and redundant manner \cite{Almheiri:2014lwa,Harlow:2016vwg}. This relation suggests that the very fabric of spacetime is not directly determined by entanglement structures, but rather by how quantum information is redundantly encoded in the underlying quantum system, positioning QEC as the fundamental language through which gravity and quantum information are understood in a unified manner.

Despite this progress, it remains unclear which aspects of standard QEC are inherently important in QG, and to what extent the connection should be understood as a precise physical principle rather than a useful analogy. In the following, we ask whether this connection can be elevated from a heuristic analogy to a precise criterion. Concretely, Section~\ref{subsec:information_recovery} focuses on how QEC-inspired toy models of black holes can be made more physical. In Section~\ref{subsec:holographic_code}, we ask which notions of QEC are actually required in QG, and whether QEC-based ideas can serve as organizing principles for low-energy theories. Addressing these questions will inevitably involve more refined discussions of QEC than hitherto known in the literature. 

\subsection{Can we incorporate more physical constraints into 
QEC toy models of QG?}\label{subsec:information_recovery}

One of the central challenges in QG is to understand black hole evaporation and the fate of the information in a black hole~\cite{Hawking:1975vcx,Page:1993wv}. Black hole information recovery can be formulated as a problem of QEC.

\newcommand{\textFourOne}{How can one refine the Hayden--Preskill toy model to incorporate more realistic black hole physics?}
\problem{\textFourOne}

The Hayden--Preskill protocol~\cite{Hayden:2007cs}, a quantum-information-theoretic approach to the black hole information paradox, has served as a canonical framework for analyzing information recovery from evaporating black holes. By modeling the black hole dynamics as a Haar-random unitary and the Hawking radiation as an accessible random subsystem~\cite{Page:1993wv}, it provides a sharp information-theoretic criterion for when information thrown into a black hole becomes decodable from the radiation.

Despite its conceptual importance, this conclusion relies on a highly simplified description of evaporation. In its standard formulation, the protocol admits an effective erasure-channel interpretation. Implicit in this description are two strong assumptions: (i) that the leakage of degrees of freedom occurs with respect to a well-defined tensor-product factorization of the Hilbert space, and (ii) that the identity of the emitted subsystem is precisely known. Under these assumptions, the recovery problem reduces to evaluating the (possibly entanglement-assisted) communication capacity of erasure channels~\cite{Bennett:1997mm,Bennett:1999hf}, rendering the problem analytically tractable.

In realistic settings, however, Hawking radiation is intrinsically quantum and probabilistic, and one cannot access information about which subsystem has been emitted. Indeed, the radiation process is not a simple deterministic peeling off of qubits from a fixed tensor structure. Moreover, beyond semiclassical approximations, even the notion of a well-defined subsystem may become ambiguous. The evaporation process is therefore unlikely to be faithfully captured by an erasure channel acting on a fixed tensor-product factor.

A more realistic model would treat the radiation process as a quantum channel compatible with semiclassical expectations for Hawking emission, potentially incorporating stochastic emission, mode-dependent couplings, and possible correlations between emissions. Such a refinement may require relaxing the naive subsystem structure in the original setup. Since recoverability thresholds, decoding complexity, and, in holographic settings, entanglement wedge reconstruction~\cite{Dong:2016eik,Jafferis:2015del} may all depend sensitively on the microscopic structure of the evaporation channel, revisiting the noise model is essential to determine which Hayden--Preskill conclusions are robust and which are artifacts of the erasure idealization.

Constructing an analytically tractable yet physically grounded information-theoretic model of black hole evaporation thus remains a central open problem at the interface of quantum information and QG.

\newcommand{\textFourTwo}{What properties of many-body Hamiltonian dynamics are necessary and sufficient for near-Haar quantum encoding?}
\problem{\textFourTwo}

Related to the above problem, a common working assumption in quantum-information-theoretic toy models of QG is that the dynamics can be approximately modeled, at least for certain purposes, by a Haar-random unitary. This assumption greatly simplifies the analysis, since Haar-random unitaries, and more generally approximate unitary 2-designs, satisfy decoupling properties that can be used to establish strong QEC performance when the dynamics is used as an encoding of quantum information~\cite{Hayden:2016pru,Dupuis:2014tkp,Dupuis:2013nyi}. The same decoupling mechanism is also central to the original Hayden--Preskill analysis, where it implies that, after sufficient scrambling and evaporation, the infalling information is no longer retained in the remaining black hole.

However, such a Haar-random description cannot literally apply to time-independent Hamiltonian systems. The evolution generated by a fixed Hamiltonian forms a one-parameter family in the unitary group and is therefore highly constrained compared to a Haar random one, which is uniformly distributed over the full unitary group. In particular, such dynamics cannot be Haar random even in the long-time limit due to the spectral nature of the one-parameter family~\cite{Roberts:2016hpo}.  

This observation leads to two related questions. First, can Haar-like dynamics be replaced by more physically natural dynamics? Second, which features of Haar-like dynamics are actually necessary for capturing the relevant properties of QG~\cite{Kim:2022pfp}? Here we focus on the former question, and the latter is discussed in Problem \problink{4}{4} in a broader context. Several works have addressed the gap between Haar-like dynamics and physical Hamiltonian dynamics~\cite{Bohdanowicz:2017atj,Cui:2025teh}. In particular, recent work has shown that ensembles of constant-local time-independent Hamiltonians cannot form accurate unitary 2-designs, while design-like behavior can emerge for suitably chosen polylog-local Hamiltonians. However, it remains unclear whether the currently known Hamiltonians achieve the level of approximation required for near-Haar quantum encoding~\cite{Dupuis:2013nyi}.

It is often conjectured, based on physical intuition and lessons from quantum chaos, that sufficiently chaotic Hamiltonians may reproduce the information-theoretic consequences derived under the Haar-random assumption. Indeed, chaotic many-body systems often display random-matrix-like spectral statistics and ETH-type behavior in simple observables~\cite{DAlessio:2015qtq}. Yet it remains unclear whether these signatures are sufficient for the much more stringent requirements of QEC. Recent numerical studies suggest that they are not: simple chaotic Hamiltonian systems can fail to realize Hayden--Preskill recovery~\cite{Nakata:2023hwg}, indicating that chaoticity alone does not automatically imply Haar-like encoding performance. 

A related caveat is that low-order Haar moments are not necessarily enough. Although Clifford or stabilizer-type ensembles furnish exact unitary or state 3-designs, their algebraic structure can still constrain mixed-state and multipartite correlations. Recent results suggest that holographic states exhibit distillable and multipartite correlations that are closer to Haar-random than to stabilizer-type structure~\cite{Li:2025nxv,Mori:2024gwe,Louisia:2025bxz}. Thus, even when a model captures some entropic or stabilizer-code-like aspects of holographic QEC, it may still fall short of the encoding properties expected from genuinely Haar-random dynamics.

This raises a fundamental question: what dynamical or spectral properties of many-body Hamiltonians are actually necessary and sufficient to achieve encoding performance comparable to Haar-random dynamics? Clarifying this issue is essential for bridging abstract quantum-information models and physically realizable dynamics. Identifying the precise classes of Hamiltonians that faithfully reproduce the QEC properties predicted under random-unitary models is therefore a central problem in understanding the information-theoretic structure of many-body Hamiltonian dynamics.

\newcommand{\textFourThree}{Can black hole dynamics be learned from Hawking radiation?}
\problem{\textFourThree}

A further compelling direction is to ask whether features of black hole dynamics can, in principle, be inferred from the radiation it emits. This question may be viewed as an inverse Hayden--Preskill problem. In the usual Hayden--Preskill setting, one assumes a known interior unitary and asks whether an unknown input state can be recovered from the outgoing radiation. Here, by contrast, the input states are used as probes, and the unknown object is the interior dynamics itself. The goal is then to infer features of this dynamics from external observations of the Hawking radiation. Such a setting is meaningful only to the extent that the infalling information leaves an operationally accessible imprint on the radiation; under this assumption, it provides a natural route toward characterizing the quantum dynamics of black holes.

From a QI perspective, this inverse problem can be formulated, at least in an idealized setting, by modeling black hole evaporation as an effective quantum channel from infalling states to outgoing radiation. In this language, the task becomes a form of quantum process tomography: one attempts to learn properties of the channel by preparing suitable input states and measuring the corresponding output radiation~\cite{Chuang:1996hw,Mohseni:2008jub}. Importantly, however, the goal need not be full reconstruction of the evaporation channel. A more physically motivated objective is to identify which structural or coarse-grained features of the underlying dynamics can be inferred from accessible radiation data.

For instance, one may ask whether the black hole dynamics exhibits properties similar to those of a Haar random unitary. A concrete way to formulate this question is through unitary $t$-designs, which quantify the extent to which a random unitary reproduces the first $t$ moments of the Haar measure. However, verifying such design properties is itself computationally challenging since deciding whether a given random unitary is a good approximation to a unitary $t$-design, or is far from a $(t+1)$-design, is computationally intractable in general~\cite{Nakata:2024tla}. This fact may motivate a more modest formulation, in which one focuses on coarse-grained diagnostics of the dynamics that are imprinted in the outgoing radiation. This viewpoint is in line with recent work on holographic detection tasks, holographic complexity, and complexity or correlator diagnostics in microscopic holographic models~\cite{Chen:2020nlj,Geloun:2023zqa,Aguilar-Gutierrez:2024nau}.

The theory of QI provides a way to quantify the difficulty of these inverse learning tasks, for example in terms of sample complexity and query complexity, which measure the number of input--output samples or controlled channel uses needed to estimate a desired feature~\cite{Arunachalam:2017fpx}. Using tools from quantum learning theory, such as query-optimal channel estimation~\cite{Mele:2025apv} and, in particular, unitary-channel estimation~\cite{Haah:2023gfp}, one may develop systematic approaches to understanding black hole dynamics from input--output relations. Shadow tomography~\cite{Aaronson:2017qlb,Huang:2020tih}, where one estimates expectations of many observables by using randomized measurements, and its extensions to channels~\cite{Kunjummen:2021anr} may also provide a complementary route to estimating selected observables without reconstructing the full evaporation channel.

This viewpoint reframes black hole physics as an operational inference problem. Instead of postulating properties of the interior, one asks what aspects of the dynamics are, in principle, learnable from radiation alone, and at what fundamental resource cost. The answer would clarify which aspects of black hole dynamics are operationally accessible. Developing such a theory would offer a new bridge between QG and quantum learning theory, grounded not in assumptions about interior structure but in what can actually be inferred from observable data.

\subsection{Does QG require full protection of quantum information?}\label{subsec:holographic_code}

While the parallels between QEC and the AdS/CFT correspondence have played a central role in uncovering emergent aspects of QG, it remains unclear which specific structural features of QEC are fundamentally required in the gravitational setting. Much of the existing intuition relies on importing the machinery of QEC into holography, yet it is not obvious whether every aspect of standard QEC theory has a direct physical counterpart in QG.

\newcommand{\textFourFour}{Can we identify specific features of QEC that are needed to describe QG?}
\problem{\textFourFour}

In QEC, quantum information is encoded into a subspace of the full physical Hilbert space.\footnote{The term `physical' used here contrasts with `logical' information we want to encode. It is different from the physical Hilbert space in gauge theory and QG, where they refer to the space of states satisfying a given gauge constraint.} This subspace, which defines a quantum error-correcting code, is referred to as the \emph{code subspace}. Operators that act nontrivially on the encoded degrees of freedom in the code space are called logical operators. Conventionally, one requires protection of the entire code subspace. All logical observables, whether local or global, should be recoverable against the prescribed class of errors. While such completeness is essential for quantum information processing, it may be unnecessarily strong from a QG perspective. In holographic systems, the physically relevant information may be encoded not in the full logical algebra but in certain subalgebras, for example, those associated with entanglement wedges~\cite{Dong:2016eik,Jafferis:2015del}. If so, a weaker notion of protection, sufficient for reconstructing operationally meaningful observables, could already capture the essential physics.

Moreover, insisting on exact protection of the full code subspace leads to additional tension with Lorentzian bulk evolution. In finite-dimensional exact codes, the Eastin--Knill theorem rules out a universal set of transversal logical gates~\cite{Eastin:2009tem}, and its covariant extensions show that continuous symmetry actions, such as time evolution, are incompatible with exact error correction unless one relaxes some aspect of the code~\cite{Faist:2019ahr}. This suggests that subalgebra QEC or approximate QEC may be more natural than exact full-subspace protection in holographic settings.\footnote{There are earlier attempts to improve tensor-network models in this direction~\cite{Qi:2018shh,Kohler:2018kqk,Apel:2021tnn,Dolev:2021ofc,Akers:2024wab,Akers:2024ixq}.}

From a quantum-information-theoretic perspective, this points toward a broader framework of relaxed QEC. Existing notions already provide intermediate regimes between the protection of the full code subspace and the complete loss of recoverability. Weak decoupling relaxes complete recovery to the preservation of pairwise distinguishability, or pairwise fidelities, of input states~\cite{Hayden:2012gjq}. Partial decoupling extends decoupling to systems with a direct-sum-product structure, making it possible to handle sector-dependent information~\cite{Wakakuwa:2019ykn}. The notion of $\alpha$-bits provides another relaxation, requiring recoverability not on the entire code subspace but on every subspace up to a prescribed size, chosen after the noisy map is fixed, with the recovery allowed to depend on the chosen subspace~\cite{Hayden:2017xed}.
It is clear from these studies that QEC properties may be graded rather than binary. In a different direction, one may also consider an extension in which the protected object is not the full algebra of logical operators on a code subspace, but rather a logical subalgebra. This framework, known as operator-algebra quantum error correction (OAQEC)~\cite{Beny:2007ewj,Pastawski:2016qrs}, provides a natural language for formulating holographic quantum error-correcting codes~\cite{Almheiri:2014lwa,Pastawski:2016qrs}.

Moreover, once one restricts attention to physically motivated error models, the relevant code parameters should be refined accordingly. For instance, if errors are confined to connected spatial regions, one is naturally led to a locality-constrained version of QEC, in which errors are localized in space rather than spread over arbitrary degrees of freedom. This is closely related to burst-error correction in coding theory, which was developed for noise that damages a localized block of a code, and to its quantum extensions for errors on neighboring qubits~\cite{Vatan:1997ys,Fan:2018qbecc}. This connection suggests a possible new direction for QEC models in which correctability depends not only on how many degrees of freedom become noisy, but also on where they are located. It also calls for new quantitative notions, including necessary and sufficient conditions for correctability against contiguous errors and suitable definitions of code distance adapted to such locality-constrained error models.

Conversely, the gravitational perspective may impose constraints in the reverse direction. It is natural to ask whether genuinely global features of quantum information, such as nonlocal logical operators, can be operationally accessed in holographic systems. This question may have direct physical significance: nonlocal logical operators that are accessible only from two asymptotic boundaries could provide an avatar of bulk information enclosed within an emergent baby universe~\cite{Mori:2025jej}. If such global information can indeed be reconstructed from boundary data, then a fully fledged QEC framework, including the protection of the entire logical algebra, may ultimately be indispensable for a complete formulation of holography.

Taken together, the key question is not whether QG is described by QEC in a literal sense, but rather which aspects of QEC are physically realized, to what degree of approximation, and under what operational constraints. Clarifying this hierarchy of QEC properties would sharpen our understanding of holography and distinguish structural necessities from convenient analogies.

\newcommand{\textFourFive}{What are possible links between Swampland conditions and holographic QEC structures?}
\problem{\textFourFive}

We now turn to a complementary perspective and discuss the implications of the QEC property for particle physics.
If (a relaxed version of) QEC is indeed a fundamental feature of QG, one may ask when and how this viewpoint fits within the traditional Wilsonian framework, in which the infrared theory of QG is described by effective field theories. Not every gravitational EFT is expected to realize the specific recoverability structure suggested by holography. The presence of such a property could serve as a sharp criterion that distinguishes consistent low-energy effective theories of QG from inconsistent ones. This idea aligns with the Swampland program~\cite{Vafa:2005ui} (see \cite{Brennan:2017rbf,Palti:2019pca,vanBeest:2021lhn,Grana:2021zvf,Agmon:2022thq} for reviews), which aims to elucidate the ``kinematics" of QG. In this sense, the QEC property would provide a new Swampland condition for low-energy theories. It is also expected to offer a fresh perspective on existing Swampland conjectures. There are several attempts in this direction, such as those in~\cite{Nakayama:2015hga,Harlow:2018tng,Montero:2018fns}.

One of the central Swampland conjectures is that the number of quantum field theories that can be consistently coupled to QG is expected to be finite~\cite{douglasprivate,Vafa:2005ui,acharya2006finite,Hamada:2021yxy}. This is one of the reasons why studying QG is not merely an ultra-high-energy issue, but is important even for low-energy physics. 
This finiteness conjecture is motivated by conjectured finiteness of the Calabi-Yau manifolds, finiteness of the flux choices~\cite{Grimm:2020cda}, and finite generalized entropy~\cite{Susskind:1994sm,Jensen:2023yxy,Kudler-Flam:2023qfl,Gesteau:2023hbq}.
Here, two (a priori) distinct notions of finiteness arise. 
\begin{enumerate}[label=\alph*)]
    \item Each bulk EFT has finite degrees of freedom. For instance, the bulk matter central charge must be finite.
    \item The number of possible quantum gravitational EFTs is finite in the low-energy theory space. The choice of couplings and their values is limited and not arbitrary.
\end{enumerate}

While such Swampland conditions have been discussed in string compactification, their consequences in information-theoretic models such as holographic QEC codes are largely unexplored.
The first key question is to understand the finiteness conjecture in the QEC context. 
a) The finiteness of each EFT degree of freedom implies that the corresponding code subspace cannot be arbitrarily large once we fix the bulk EFT cutoff.
b) The finiteness of the number of quantum gravitational EFTs implies that there should be only finitely many independent QEC codes given a holographic quantum gravity with a fixed boundary condition.
In principle, the gravitational path integral may include infinitely many saddles of the Einstein equation. Around each saddle, one can define a perturbative quantum gravity theory, and each such perturbative theory may be interpreted as specifying a corresponding holographic QEC code~\cite{Jafferis:2015del,Chen:2019gbt}. However, the finiteness conjecture implies that only finitely many of these saddles should give rise to consistent quantum gravitational EFTs. Consequently, only a finite number of holographic QEC codes can provide valid holographic encodings. This may be interpreted as a higher-order equivalence among QEC codes themselves. Different QEC codes could be equivalent for a specific class of states, namely, \emph{physical} states that satisfy the gravitational constraints.

Assuming that the finiteness conjecture holds in the QEC context, we can then ask the more refined question raised in point b).
In addition to the finiteness of possible choices for couplings and masses in bulk EFTs, do these values obey specific patterns? 
One of the Swampland conjectures relevant to this issue is the weak gravity conjecture~\cite{Arkani-Hamed:2006emk} (WGC), which states that gravity should be the weakest force. In a theory with a $U(1)$ gauge force, there should exist a charged state whose gauge repulsion is strong enough compared with its gravitational attraction.
In the QEC language, this again corresponds to specifying the way to encode the code subspace into the entire Hilbert space. 
It is then interesting to ask whether such a condition for WGC is naturally realized in the holographic QEC framework, or whether it imposes any additional constraints.

The species bound~\cite{Dvali:2007hz,vandeHeisteeg:2022btw} provides another possible Swampland constraint on holographic quantum error correction. It states that, in the presence of $N$ light particle species, the gravitational cutoff is lowered parametrically as
\[
\Lambda_{\rm grav}\sim \frac{M_{\rm Pl}}{N^{\frac{1}{d-2}}},
\]
so that the bulk EFT breaks down at a lower energy scale than one would infer from the Planck scale alone. This observation may have important consequences for holographic QEC, especially in connection with the black hole information paradox.

In the original Ryu-Takayanagi prescription, the entropy is determined by minimizing the area term $A/4G_N$. The island formula, however, makes clear that the bulk matter contribution can compete with the area term in the generalized entropy~\cite{Penington:2019npb,Almheiri:2019psf,Almheiri:2019hni,Almheiri:2019qdq,Penington:2019kki2,Goto:2020wnk}:
\[
S_R=\min \underset{I}{\mathrm{ext}}
\left[
\frac{\mathrm{Area}(\partial I)}{4G_N}
+S_{\rm bulk}(R\cup I)
\right].
\]
After the Page time, the saddle that minimizes only the area term is no longer the dominant one; instead, the dominant QES includes an island $I$ containing part of the black hole interior. A large number of light species can enhance the bulk entropy contribution, since $S_{\rm bulk}$ typically receives contributions from all light fields. Thus, even when each species has a small gravitational effect, their collective contribution may alter the competition between the area and matter terms and can potentially make the island saddle dominant at an earlier time.

From the viewpoint of holographic QEC, this suggests that the Species Bound may affect the regime in which black hole interior operators become reconstructible from the radiation. In particular, by enhancing the matter contribution to the generalized entropy and lowering the gravitational cutoff, many light species may shift the threshold for entanglement-wedge reconstruction. This provides a possible bridge between Swampland constraints, the black hole information paradox, and the operational structure of holographic QEC.

\newcommand{\textFourSix}{What is the holographic counterpart of the classical--quantum correspondence of QEC?}
\problem{\textFourSix}

Having discussed possible bulk EFT constraints on admissible holographic codes, we now turn to a different but equally basic aspect of their structure: the nature of the information they protect. 
If QEC is indeed a structural feature of holography, a natural question from the QI perspective is how quantum the protected information must be. For a given code, protection of quantum information is closely related to protection of two associated kinds of classical information. This relation leads to the classical--quantum correspondence: quantum recoverability can be characterized in terms of classical recoverability in two complementary bases~\cite{Renes:2016olj,Nakata:2022xny}. Despite this structural correspondence, their ultimate communication capabilities differ, as reflected in the Holevo--Schumacher--Westmoreland theorem~\cite{Holevo:1996yz,Schumacher:1997cde} for classical communication and the Lloyd--Shor--Devetak theorem~\cite{Lloyd:1996at,Shor:2002qcc,Devetak:2004erm} for quantum communication. This difference suggests a hierarchy in QEC according to the nature of the information being protected.

This hierarchy suggests that holography itself may admit different levels of protection. Rather than assuming that the boundary theory encodes the full noncommutative bulk algebra, one may ask whether it robustly protects only a restricted set of mutually commuting bulk observables, corresponding to classical information in a single basis. In such a ``classical holographic'' regime, classical geometric features of the bulk could remain stable even if generic bulk observables are not fully protected.

The classical--quantum correspondence then raises a natural and more ambitious possibility. Since quantum information is equivalent to two kinds of classical information in complementary bases, one may speculate that ``quantum holography'' could emerge from two classical holographic encodings. If a boundary theory were to stably encode two noncommuting sets of bulk observables, each individually behaving as a classical holographic code, their joint protection could amount to full quantum protection of the bulk degrees of freedom. In this sense, quantum holography might be understood as arising from two complementary classical holographic structures.

This perspective aligns naturally with OAQEC. Within OAQEC, the distinction between classical and quantum protection can be captured algebraically: protecting a commutative subalgebra corresponds to classical protection, whereas protecting a sufficiently rich noncommutative algebra corresponds to quantum protection. From this standpoint, a central question for QG becomes: which collections of bulk subalgebras are simultaneously protected holographically, and how incompatible must they be in order to give rise to fully quantum behavior?

Identifying the holographic counterpart of the classical--quantum correspondence would offer a fresh perspective on holography. In particular, exploring ``classical holographic'' regimes could reveal intermediate layers of encoding that are not directly visible within the standard full-QEC interpretation. 
Such an investigation may help distinguish which aspects of bulk geometry rely on genuinely quantum protection from those that already arise at a classical level.

\section{Infinite dimensionality}\label{sec:infinite_dim}
String theory and QFT involve operators acting on inherently infinite-dimensional Hilbert spaces. Infinite-dimensionality of Hilbert spaces is also relevant in the context of the thermodynamic limit of finite-size systems, as well as continuous-variable (CV) systems such as photonic systems. A potential mismatch arises from the fact that many foundational techniques in QI are developed under the assumption that the underlying Hilbert spaces are finite-dimensional. Consequently, the conclusions drawn from such techniques often do not automatically extend to infinite-dimensional settings, limiting the theoretical justification for applying entropic quantities in these physically relevant application domains.

\subsection{Can we establish the theory of QI in infinite-dimensional settings, in particular QFTs?}

The first obstacle against extending the theory of QI to continuum theories is the fact that the Hilbert space is not factorizable. This happens because of the short-distance behaviour, where there are more and more entangled modes closer and closer to the boundary between subregions. Part of this problem was successfully solved using the celebrated modular theory of von Neumann algebras by Tomita and Takesaki, where relative entropy can be defined, for example \cite{Takesaki:1970aki}. However, the definition of entanglement entropy is more subtle and requires regularization, only after which one obtains a factorized Hilbert space that approximates the original continuum one~\cite{Hellerman:2021fla}. Even then, such Hilbert spaces are infinite-dimensional, and we are required to have a general theory of quantum information in infinite-dimensional Hilbert spaces if we ever hope to study QI in continuum theories.

\newcommand{\textFiveOne}{What are the appropriate mathematical toolsets and arguments for canonically extending quantum information theory to infinite-dimensional settings, such as those encountered in QFT?}
\problem{\textFiveOne}

In general, results established in QIT for finite-dimensional settings do not necessarily generalize to infinite-dimensional systems. For example, a fundamental task in entanglement theory is entanglement distillation, which aims to convert noisy entangled states into pure maximally entangled states; its optimal rate, known as distillable entanglement, provides an operationally meaningful quantification of entanglement~\cite{Bennett:1995tk}. However, characterizing this quantity remains notoriously difficult, even for mixed states in finite-dimensional systems. The reverse task, known as entanglement dilution, involves transforming pure maximally entangled states into a given noisy entangled state, which is equally fundamental in entanglement theory~\cite{Bennett:1995tk}. Its optimal rate is referred to as the entanglement cost. A cornerstone of entanglement theory is the characterization of entanglement cost via the regularized entanglement of formation, originally established in~\cite{hayden2001asymptotic}. However, the analysis in~\cite{hayden2001asymptotic} relies on the assumption that the quantum systems manipulated by LOCC are represented by finite-dimensional Hilbert spaces. 
In principle, if LOCC were allowed to operate on infinite-dimensional systems, it could become strictly more powerful, potentially altering the optimal rates, even when the target states are finite-dimensional; moreover, when the target states are infinite-dimensional, the protocol in~\cite{hayden2001asymptotic} no longer applies.

A particularly sharp manifestation of this subtlety is provided by entanglement embezzlement~\cite{vanDam:2002jri}. 
Embezzling states allow Alice and Bob, given a suitable shared entangled catalyst, to approximately generate an arbitrary bipartite pure entangled state using local unitaries alone and without classical communication, while leaving the catalyst nearly unchanged.
Recent work has shown that this phenomenon becomes much more rigid and universal in the von Neumann algebraic setting: type III factors admit embezzling states, type III$_1$ factors are universal embezzlers, and relativistic quantum fields therefore provide natural examples of universal entanglement embezzlers~\cite{vanLuijk:2024ygl,vanLuijk:2024nnx}. Closely related results extend the LOCC theory of pure-state transformations to arbitrary factors, showing in particular that type III factors allow arbitrary-precision transitions between pure states and that, for type III$_1$ factors, such transitions can even be achieved without classical communication~\cite{vanLuijk:2024cop}.
These results illustrate why the infinite-dimensional limit of finite-dimensional quantum information theory cannot be treated merely as formal limits: auxiliary infinite-dimensional degrees of freedom may effectively act as unbounded reservoirs of entanglement.\footnote{
See also~\cite{vanLuijk:2024rby} for universal embezzlement in critical one-dimensional free-fermionic systems, 
\cite{vanLuijk:2024oao} for multipartite embezzlement, 
\cite{Schwartzman:2024zim} for complexity-theoretic obstructions to perfect embezzlement, 
and~\cite{vanLuijk:2025ufz,vanLuijk:2025gkb} for broader perspectives on von Neumann algebraic quantum information theory and large-scale entanglement in many-body systems.}

Recently, \cite{yamasaki2025entanglement} resolved an important instance of this broader issue within the Hilbert-space formulation of QI by proving that, for quantum states on infinite-dimensional separable Hilbert spaces satisfying suitable energy constraints, the entanglement cost equals the regularized entanglement of formation.
This proof requires more advanced techniques than those used in its finite-dimensional counterpart in~\cite{hayden2001asymptotic}, such as semi-continuity bounds for the entanglement of formation in infinite dimensions~\cite{Shirokov:2022vyl} and careful handling of the mathematical subtleties that arise in the infinite-dimensional setting.
Of course, this result does not by itself close the full conceptual and technical gap between finite- and infinite-dimensional quantum mechanics; many key tools in QI inherently depend on finite-dimensionality. It is essential to develop further toolkits and canonical arguments to investigate the theory of QI in infinite-dimensional settings, in a manner more compatible with QFT frameworks.

A natural expectation is that any infinite-dimensional setting arises as a suitable limit of finite-dimensional ones. 
This intuition can be given a precise mathematical formulation as follows.
In the theory of von Neumann algebras, the hyperfinite type II$_1$ factor $R$ is defined as a suitable limit (weak closure of an inductive limit) of the $n\times n$ matrix algebras, where the traces are scaled so that the trace of the identity operator is normalized to be one. We can also consider a (not necessarily hyperfinite) type II$_1$ factor, which is infinite-dimensional but is equipped with a well-defined trace canonically normalized on the identity. Now, Connes's embedding problem (CEP) \cite{MR454659} asks if any type II$_1$ factor can be regarded as a limit of the matrix algebra similar to the hyperfinite case; slightly more precisely, if any separable type II$_1$ factor can be embedded into an ultrapower $R^{\omega}$ of the hyperfinite II$_1$ factor $R$, where $\omega$ is a free (non-principal) ultrafilter on $\mathbb{N}$. 

Despite Connes's expectation, the CEP has recently been shown to fail \cite{Ji:2020apg}, as a corollary of the complexity-theoretic theorem ``$\mathrm{MIP}^* = \mathrm{RE}$'' \cite{Ji:2020apg} (see \cite{Goldbring:2021prg} for a pedagogical overview).\footnote{This result provides a striking contrast to the aforementioned operational limitations of an observer. While an observer's access is typically hindered by causal or computational constraints, the $\mathrm{MIP}^* = \mathrm{RE}$ theorem demonstrates that if provers share unbounded entanglement, their collective verification power explodes to encompass the entire class of recursively enumerable languages, including the halting problem. This suggests that in the infinite-dimensional Hilbert spaces of QFT, the very definition of ``information restriction" undergoes a radical transformation through nonlocal correlations.}
The CEP is moreover known to be equivalent to several reformulations, notably Kirchberg's QWEP conjecture and Tsirelson's problem \cite{MR1218321,Fritz:2012kbj,Junge:2010cuf,ozawa2012connes}---Tsirelson's problem, for example, asks 
whether quantum correlations realized by commuting observables on a single joint Hilbert space can always be reproduced by observables acting on separate tensor factors.
One should also note that type II$_{1}$ factors can serve as building blocks for the type III$_1$ factors, via tensoring with the bounded operators $B(\mathcal{H})$ on an infinite-dimensional Hilbert space $\mathcal{H}$ and a trace-scaling crossed product \cite{MR438149,MR341115,landstad1979duality}. The type III$_1$ factor is of particular importance since the local algebras of observables in QFT, including bulk EFT, are in general believed to be described by a type III$_1$ factor. Taken together, the discussions above highlight fundamental subtleties in infinite-dimensional settings.


\subsection*{Acknowledgements}

This project emerged out of a workshop ``Next-Generation Workshop: Toward the Future of Strings and Quantum Information,'' November 7-9, 2025, which was partly supported by IBM-UTokyo-sponsored research (PI: M.Y.). 
K.G. was supported by JSPS KAKENHI Grant Nos. 24K17048 and 22J00663.
Y.H. was supported in part by JSPS KAKENHI Grant Nos. JP24H00976, JP24K07035, JP24KF0167, and by JST BOOST Program Japan Grant No. JPMJBY25E1.
K.K. acknowledges support from MEXT-JSPS Grant No. 24H00829; from JSPS KAKENHI Grant No. 23K17668; from MEXT Quantum Leap Flagship Program (MEXT QLEAP) Grant No. JPMXS0120319794.
T.M. was supported by JSPS KAKENHI Grant Nos. 23KJ1154, 24K17047, and by the RIKEN TRIP-CoRe initiative (RIKEN Quantum). 
Y.~N. was supported by JST PRESTO Grant No. JPMJPR2456 and JST CREST Grant No. JPMJCR23I3.
H.Y. was supported by JST PRESTO Grant No. JPMJPR23FC, JST CREST Grant No. JPMJCR25I5, JST Moonshot R\&D Grant No. JPMJMS256J, and Faculty Research Funding from Google Quantum AI\@.
M.Y. was supported in part by the World Premier International Research Center Initiative (WPI), MEXT, Japan; by the JSPS KAKENHI Grant No. 23K25865; by JST, Japan (CREST Grant No.\ JP-MJCR26XA, Moonshot R\& D Grant No.\ JPMJMS2061); and by the IBM-UTokyo-sponsored research.
T.Y. was supported in part by JSPS KAKENHI Grant Nos. JP22H05115 and 25K17390.

\appendix

\newcommand{\problemitem}[4]{%
    \gdef\currentproblemtarget{prob-#1-#2}%
    \item #3%
    \if\relax\detokenize{#4}\relax
    \else
    \fi
}

\section{List of problems}\label{sec:list}
This appendix summarizes the problems discussed throughout the paper.
Each entry is hyperlinked to the corresponding problem box in the main text.

\setcounter{subsection}{1}
\subsection{Operational characterization}

\begin{enumerate}[label=\protect\problemlistlabel]
\problemitem{2}{1}{\textTwoOne}{\problink{2}{2}, \problink{3}{1}, \problink{4}{4}}
\problemitem{2}{2}{\textTwoTwo}{\problink{2}{1}, \problink{3}{3}, \problink{4}{1}, \problink{4}{4}}
\problemitem{2}{3}{\textTwoThree}{\problink{2}{1}, \problink{2}{4}, \problink{4}{4}, \problink{5}{1}}
\problemitem{2}{4}{\textTwoFour}{\problink{2}{3}, \problink{3}{5}, \problink{4}{1}, \problink{4}{3}}
\problemitem{2}{5}{\textTwoFive}{\problink{2}{6}, \problink{2}{7}, \problink{2}{8}, \problink{5}{1}}
\problemitem{2}{6}{\textTwoSix}{\problink{2}{5}, \problink{2}{7}, \problink{2}{8}}
\problemitem{2}{7}{\textTwoSeven}{\problink{2}{5}, \problink{2}{6}, \problink{2}{8}}
\problemitem{2}{8}{\textTwoEight}{\problink{2}{5}, \problink{2}{6}, \problink{2}{7}, \problink{4}{4}, \problink{4}{5}, \problink{4}{6}}
\end{enumerate}

\subsection{Observer}

\begin{enumerate}[label=\protect\problemlistlabel]
\problemitem{3}{1}{\textThreeOne}{\problink{2}{1}, \problink{3}{2}, \problink{3}{3}}
\problemitem{3}{2}{\textThreeTwo}{\problink{3}{1}, \problink{3}{3}, \problink{3}{4}}
\problemitem{3}{3}{\textThreeThree}{\problink{2}{2}, \problink{3}{1}, \problink{3}{2}, \problink{3}{4}}
\problemitem{3}{4}{\textThreeFour}{\problink{2}{4}, \problink{3}{2}, \problink{3}{3}, \problink{3}{5}}
\problemitem{3}{5}{\textThreeFive}{\problink{2}{4}, \problink{3}{3}, \problink{3}{4}, \problink{5}{1}}
\end{enumerate}

\subsection{Quantum error correction}

\begin{enumerate}[label=\protect\problemlistlabel]
\problemitem{4}{1}{\textFourOne}{\problink{2}{2}, \problink{2}{4}, \problink{4}{2}, \problink{4}{3}}
\problemitem{4}{2}{\textFourTwo}{\problink{4}{1}, \problink{4}{3}, \problink{4}{4}}
\problemitem{4}{3}{\textFourThree}{\problink{2}{4}, \problink{4}{1}, \problink{4}{2}}
\problemitem{4}{4}{\textFourFour}{\problink{2}{1}, \problink{2}{2}, \problink{2}{8}, \problink{4}{2}, \problink{4}{6}}
\problemitem{4}{5}{\textFourFive}{\problink{2}{8}, \problink{4}{4}, \problink{4}{6}}
\problemitem{4}{6}{\textFourSix}{\problink{2}{8}, \problink{4}{4}, \problink{4}{5}}
\end{enumerate}

\subsection{Infinite dimensionality}

\begin{enumerate}[label=\protect\problemlistlabel]
\problemitem{5}{1}{\textFiveOne}{\problink{2}{3}, \problink{2}{5}, \problink{3}{5}, \problink{4}{4}}
\end{enumerate}


\mciteSetMidEndSepPunct{}{\ifmciteBstWouldAddEndPunct.\else\fi}{\relax}
\small 
\setstretch{0.9} 
\setlength{\bibsep}{0pt plus 0.3ex} 
\bibliographystyle{JHEP}
\bibliography{ref.bib}

\end{document}